\documentclass[twocolumn, eqsecnum, aps ]{revtex4}
\usepackage{graphicx}
\usepackage{dcolumn}

\begin{document}

\title{{\it Ab initio} determination of the lifetime of the $6^2P_{3/2}$ state for $^{207}Pb^+$ by relativistic many-body theory} 
\vspace{0.5cm}

\author{Bijaya K. Sahoo, $^\ddagger$ Sonjoy Majumder, Rajat K. Chaudhuri, B.P. Das\\ and \\$ ^{\dagger}$Debashis Mukherjee\\
\vskip0.3cm
Non-Accelerator Particle Physics Group\\Indian Institute of Astrophysics, Bangalore-34, India\\
$^\ddagger$ Institut f\"ur Theoretische Chemie, Technical University of M\"unich,\\85747,Garching, Germany\\
$ ^{\dagger}$Indian Association for Cultivation of Science, Calcutta-700 032, India} 
\date{Received date; Accepted date}
\vskip1.3cm
\begin{abstract}
\noindent
Relativistic coupled-cluster(RCC) theory has been employed to calculate the lifetime of the $6 ^2P_{3/2}$ state of single ionized lead($^{207}Pb$) to an accuracy of 3\% and compared with the corresponding value obtained using second order relativistic many-body perturbation theory(RMBPT). This is one of the very few applications of this theory to excited state properties of heavy atomic systems. Contributions from the different electron correlation effects are given explicitly.
\end{abstract} 
\maketitle

\section{Introduction}
\noindent
Trapped and laser cooled ions are excellent candidates for high precision measurements\cite{pradip01,wineland01}. $^{207}Pb^+$ is the heaviest atomic ion to be trapped and cooled to date\cite{strumia,werth01}. The lifetime of the $6p_{3/2} (6 ^2P_{3/2})$ state of this ion has been measured to an accuracy of 2\%\cite{werth02}. The transition from this excited state to the ground state $6p_{1/2} (6 ^2P_{1/2}$) is predominantly magnetic dipole(M1) in character, but there is a small electric quadrupole(E2) component as well. The M1 and E2 transition rates depend on the inverse cubic and quintic powers of the transition wavelength respectively. The lifetime of the $6 ^2P_{3/2}$ for $Pb^+$ must therefore be calculated by an accurate relativistic many-body method. Indeed a comparision of the measured and calculated values of the lifetime for this state would constitute a stringent test of the theoretical approach on which the calculation is based. In addition, a precise knowledge of this lifetime is useful in determining the abundance of Pb in the planetary nebula NGC 7027\cite{burris,pequignot}. The work in this paper uses the non-linear RCC theory to calculate this lifetime. This theory has been successfully applied earlier to different atomic systems\cite{geetha01,sahoo01,sahoo02,sonjoy01}, but it has seldom been used to study excited state properties for heavy atoms or ions. This is the first accurate calculation of the lifetime of the $6p ^2P_{3/2}$ state of $Pb^+$ to our knowledge. We compare our RCC results for the excited energy and lifetime with those obtained from second order RMBPT(2).

\section{THEORY}
\noindent
The one-electron reduced matrix elements of M1 and E2 operators are given by\cite{berestetski71}
\begin{eqnarray}
\langle j_f || q_m^{(1)} || j_i \rangle 
= \langle j_f || C_m^{(1)} || j_i  \rangle \frac {6}{\alpha k} \frac {(\kappa_f + \kappa_i)}{2} \nonumber \\ 
\times {\int dr j_1(kr)(P_{\kappa_f}Q_{\kappa_i}+Q_{\kappa_f}P_{\kappa_i})},
\end{eqnarray}
and
\begin{eqnarray}
\langle j_f || q_m^{(2)} || j_i \rangle
=  \langle j_f || C_m^{(2)} || j_i  \rangle  \frac {15}{k^2}\hspace*{2cm} \nonumber \\
\times \int dr \hspace*{0.2cm} \{ j_2(kr) \hspace*{0.2cm}
(P_{\kappa_f} P_{\kappa_i}+Q_{\kappa_f}Q_{\kappa_i}) \nonumber \\
 + j_3(kr) \frac {(
j_f - j_i - 1 )}{3}
 (P_{\kappa_f}Q_{\kappa_i} + Q_{\kappa_f}P_{\kappa_i})\}
\end{eqnarray}
\noindent
respectively, where $j_i$ and $\kappa_i$ are the total angular momentum and relativistic angular momentum($\kappa_i=\pm(j_i+\frac {1}{2}$)) quantum numbers respectively of the {\it i'th} electron orbital and $k=\omega \alpha$, where $\omega = \epsilon_f - \epsilon_i$, is the photon energy of the transition, $\alpha$ is the fine-structure constant. 
We use atomic units ($\hbar = m_e = |e| = 1$) in this paper. The quantity $C_m^{(1)}(\hat r)$ is the Racah tensor and $j_l(kr)$ is the spherical Bessel function of order {\it l}. $P_{\kappa_i}$ and $Q_{\kappa_i}$ are the large and small radial components of the Dirac-Fock $i^{th}$ single particle wave functions.\\

\noindent
The angular factor is given by
\begin{eqnarray}
<j_f||C_m^{(l)}||j_i> = (-1)^{(j_f+1/2)} \sqrt{j_f+1/2} \sqrt{j_i+1/2}\nonumber \\
                                  \times \left ( \matrix{
                                          j_f & l & j_i \cr
                                          1/2 & 0 & -1/2 \cr
                                          }
                                         \right ) 
\end{eqnarray}
\noindent
The M1 and E2 transition probabilities $A_{FI}(s^{-1})$ can be expressed in terms of the line strength $S_{FI}$ which is the square of the sum of the single particle transition matrix elements given by eq(2.1) and eq(2.2) for the appropriate transitions and wavelength $\lambda(A^0)$ as\cite{drake,sobelman}
\begin{equation}
A_{FI}^{M1} = \frac {2.6973 \times 10^{13}}{[J_I]\lambda^3} S_{FI}^{M1}
\end{equation}
and
\begin{equation}
A_{FI}^{E2} = \frac {1.1199 \times 10^{18}}{[J_I]\lambda^5} S_{FI}^{E2}
\end{equation}
where J$_I$ is the degeneracy of the initial metastable state which is equal to 3/2 for the present calculation.

\section{METHOD OF CALCULATION : Relativistic Coupled Cluster Theory}
\noindent
The relativistic Dirac-Coulomb atomic Hamiltonian is given by
\begin{equation}
H = \sum_{j} c \vec {\bf\alpha}.\vec {\bf p}_{j} + (\beta -1) c^{2} + V_{nuc}(r_j) +
\sum_{j<l} \frac {1}{r_{jl}}
\end{equation}
where $\vec \alpha$ and $\beta$ are the usual Dirac matrices and $V_{nuc}(r_j)$ is the potential at the site of the $j^{th}$ electron due to the atomic nucleus.
The energy eigen values are scaled with respect to the rest mass energy of the electron. We first solve the relativistic Hartree-Fock(Dirac-Fock) equations to obtain the single particle orbitals and their energies.
$$H_{DF} = \sum_{j} c \vec {\bf\alpha}.\vec {\bf p}_{j} + (\beta -1)c^{2} + V_{nuc}(r_j) + U_{DF}(r_j)$$
The residual Coulomb interaction is given by 
\begin{equation}
V_{es} = \sum_{j<l} \frac {1}{r_{jl}} - \sum_{j} U_{DF}(r_j)
\end{equation}
The single particle orbitals are obtained by solving the following equation self-consistently
\begin{equation}
(t_j + U_{DF}(r_j) ) |\phi_j\rangle = \epsilon_j|\phi_j\rangle
\end{equation} 
where
\begin{eqnarray}
t_j &=& c \vec \alpha \cdot \vec {\bf p}_j + (\beta - 1) c^2 +V_{nuc}(r_j)\nonumber
\end{eqnarray}
and
\begin{eqnarray}
U_{DF} |\phi_j\rangle &=& \sum \langle \phi_a|\frac{1}{r_{jl}}|\phi_a\rangle |\phi_j\rangle - \langle \phi_a|\frac{1}{r_{jl}}|\phi_j\rangle |\phi_a\rangle\nonumber
\end{eqnarray}
The single particle relativistic orbitals can be expressed as
$$
|\phi_j(r)\rangle = \frac {1}{r} \left (
                        \matrix {
                         P_j(r) |\chi_{\kappa_j m_j} \rangle \cr
                         Q_j(r) |\chi_{-\kappa_j m_j} \rangle \cr 
                         }
                        \right )
$$
where $P_j(r)$ and $Q_j(r)$ are the radial part of the large and small components respectively and $|\chi_{\kappa_j m_j}\rangle$ and $|\chi_{-\kappa_j m_j}\rangle$ are their respective spin angular momentum components. $\epsilon_j$'s are the single particle energies.\\

\noindent
We have employed the RCC to incorporate correlation effects among electrons due to the residual Coulomb interaction. In this approach the exact atomic wavefunction for the closed-shell system can be expressed as\cite{lindgren01} 
\begin{equation}
|\Psi_{CC}\rangle = e^T |\Phi\rangle
\end{equation}
where T is the core electron excitation operator and $|\Phi\rangle$ is the above closed-shell determinantal state built out of the Dirac-Fock single particle orbitals.\\

\noindent
In the closed-shell coupled-cluster theory one starts with the equation
\begin{equation}
H e^T |\Phi\rangle = E e^T |\Phi\rangle
\end{equation}
The energy and amplitude determining equations are
\begin{equation}
\langle \Phi^K |\bar{H}|\Phi \rangle = E \delta_{K,0} 
\end{equation}
where $\bar{H} = e^{-T} H e^{T}$, $|\Phi^K\rangle$ is a determinantal state with K = 0,1,2.... representing the reference state and excited determinantal states.
We have considered all possible non-linear terms in T- operator for its amplitude determining equations. \\ 

\noindent
Goldstone\cite{lindgren01,szabo01,bartlett} and angular momentum diagrammatic\cite{lindgren01,edmonds01} techniques are used for evaluating different radial integrals and angular factors. The normal ordered Hamiltonian is defined as
\begin{equation}
H_N \equiv H - \langle \Phi |H| \Phi \rangle =  H - E_{DF},
\end{equation}
where $E_{DF}=\langle \Phi|H|\Phi\rangle$.\\

\noindent
We have truncated our wavefunction expansion at the level of singles and doubles(CCSD) and all possible non-linear terms have been included in the above equation.\\ 

\noindent
The ground state of $Pb^+$ contains only one valence electron; namely the $6p_{1/2}$ orbital. One way 
to evaluate the ground state energy of $Pb^+$ is to first compute the wavefunctions for
the closed shell system $Pb^{++}$ using the above closed shell CC approach and then append a valence electron 
 ($6p_{1/2}$ orbital) using the open shell CC (OSCC) method
 as follows. The same procedure has been followed to obtain the excited $6p_{3/2}$ state.\\
  
\noindent
The new reference state of the open-shell system with one valence electron {\it v} can be expressed as\cite{debasish} 
\begin{equation}
|\Phi_v\rangle\ \equiv\ a_v^{\dag}|\Phi\rangle
\end{equation}
where $a_v^{\dag}$ is the particle creation operator. The exact atomic states are defined now, using the Fock-space OSCC method, as\cite{lindgren01,debasish} 
\begin{equation}
|\Psi_v\rangle = e^T\{e^{S_v}\}|\Phi_v\rangle
\end{equation}
where $S_v$ is the valence excitation operator. Since the system under consideration has only one valence electron, the S- operator exponetial series naturaly truncates at linear term, i.e. the open-shell wavefunction has the form
\begin{equation}
|\Psi_v\rangle = e^T\{1+S_v\}|\Phi_v\rangle
\end{equation}
where
\begin{eqnarray}
S_v\ =\ S_{v1} + S_{v2} = \sum_{p \ne v}a_p^+a_v s_v^p + \frac {1}{2}\sum_{bpq}a_p^+a_q^+a_ba_v s^{pq}_{vb} \nonumber
\end{eqnarray}
and
\begin{eqnarray}
 S_{v1} = \sum_{p \ne v} a_p^+a_v s_v^p \nonumber \\
 S_{v2} = \frac {1}{2}\sum_{bpq}a_p^+a_q^+a_ba_v s^{pq}_{vb}
\end{eqnarray}
with $s_v^p$ and $s^{pq}_{vb}$ are the cluster amplitudes corresponding to single and double excitations involving the valence electron.\\

\noindent
In the next step, we include approximate triple excitations by contracting the two-body operator($V_{es}$) and the double excitation operators($T_2,S_{v2}$) in the following way\cite{kaldor01,kaldor02} 
\begin{equation}
S_{vbc}^{pqr}\ =\ \frac{\widehat{V_{es}T_2}+\widehat{V_{es}S_{v2}}}{\epsilon_v+\epsilon_b+\epsilon_c-\epsilon_p
-\epsilon_q-\epsilon_r}
\end{equation}
where $\epsilon_i$ is the orbital energy of the {\it i'th} orbital. Note that we use notations a,b,c..., p,q,... and i,j,... for the core, particle and generic orbitals respectively.\\

\noindent
The wavefunction in the framework of many-body perturbation theory(MBPT) can be written as
\begin{equation}
|\Psi_{MBPT}\rangle = |\Phi_v\rangle + |\Phi^{(1)}\rangle + |\Phi^{(2)}\rangle + |\Phi^{(3)}\rangle + ...
\end{equation} 
where $|\Phi^{(n)}\rangle$ is the $n^{th}$ order correction to the wavefunction  $|\Phi\rangle$. Each order corrected wavefunction is a linear combination of all excited determinantal states with respect to $|\Phi\rangle$. Gathering excitations of the same order together from each of the corrected wavefunction we can rewrite the wavefunction of the many-body system as

\noindent
\begin{eqnarray}
|\Psi_{MBPT}\rangle = |\Phi_v\rangle + \sum_{ap} (C_a^p(1,1)+C_a^p(2,1)+..) |\Phi_a^p\rangle \nonumber \\
= |\Phi_v\rangle + \sum_{p \ne v} (C_v^p(1,1)+C_v^p(2,1)+..) |\Phi_v^p\rangle \nonumber \\
+ \frac {1}{2}\sum_{abpq} (C_{ab}^{pq}(1,2)+C_{ab}^{pq}(2,2)+..) |\Phi_{ab}^{pq}\rangle + ... \nonumber\\
+ \frac {1}{2}\sum_{bpq} (C_{vb}^{pq}(1,2)+C_{vb}^{pq}(2,2)+..) |\Phi_{vb}^{pq}\rangle + ... \nonumber\\
= |\Phi_v\rangle + \sum_{ap} (C_a^p(1,1) + C_a^p(2,1)+....) a_p^{\dagger}a_a |\Phi\rangle \nonumber \\
 + \sum_{p \ne v} (C_v^p(1,1) + C_v^p(2,1)+....) a_p^{\dagger}a_v |\Phi\rangle \nonumber \\
+ \frac {1}{2}\sum_{abpq} (C_{ab}^{pq}(1,2)+C_{ab}^{pq}(2,2)+..) a_p^{\dagger}a_q^{\dagger}a_ba_a | \Phi\rangle + ....\nonumber \\
+ \frac {1}{2}\sum_{bpq} (C_{vb}^{pq}(1,2)+C_{vb}^{pq}(2,2)+..) a_p^{\dagger}a_q^{\dagger}a_ba_v | \Phi_v\rangle + ....\nonumber \\
= |\Phi_v\rangle + \sum_{ap} t_a^p a_p^{\dagger}a_a |\Phi \rangle
+ \frac {1}{2}\sum_{abpq} t_{ab}^{pq} a_p^{\dagger}
a_q^{\dagger}a_ba_a |\Phi\rangle + ....\nonumber \\
 + \sum_{p \ne v} s_v^p a_p^{\dagger}a_v |\Phi_v \rangle
+ \frac {1}{2}\sum_{bpq} s_{vb}^{pq} a_p^{\dagger}
a_q^{\dagger}a_ba_v |\Phi_v \rangle + ....\nonumber \\
+ \frac {1}{2}\sum_{abpq} t_a^p t_b^q a_p^{\dagger}
a_q^{\dagger}a_ba_a \Phi\rangle + ....\nonumber \\
= (1+ T_1 + T_2 + \frac {1}{2} T_1^2...) |\Phi_v\rangle + S_{1v} |\Phi_v \rangle + S_{2v} |\Phi_v \rangle \nonumber \\
= e^T \{1 + S_v\} |\Phi_v \rangle \hspace*{4.9cm} \nonumber \\
\equiv |\Psi_{CC}\rangle \hspace*{6.3cm}
\end{eqnarray}
where 
\begin{eqnarray}
t_a^p = C_a^p(1,1) + C_a^p(2,1) + .... \nonumber \\
s_v^p = C_v^p(1,1) + C_v^p(2,1) + .... \nonumber \\
t_{ab}^{pq} + t_a^p t_b^q = C_{ab}^{pq}(1,2) + C_{ab}^{pq}(2,2) + .... \nonumber \\
s_{vb}^{pq} = C_{vb}^{pq}(1,2) + C_{vb}^{pq}(2,2) + .... \nonumber \\
T_1 = \sum_{ap} t_a^p a_p^{\dagger}a_a \nonumber \\
S_{1v} = \sum_{p \ne v} s_v^p a_p^{\dagger}a_a \nonumber \\
T_2 = \frac {1}{2}\sum_{abpq} t_{ab}^{pq} a_p^{\dagger}a_q^{\dagger}a_ba_a \nonumber \\
S_{2v} = \frac {1}{2}\sum_{bpq} s_{vb}^{pq} a_p^{\dagger}a_q^{\dagger}a_ba_v \nonumber
\end{eqnarray}
and
\begin{eqnarray}
T = T_1 + T_2 + ...\nonumber \\
= \sum_{ap} t_a^p a_p^{\dagger}a_a + \frac {1}{2}\sum_{abpq} t^{pq}_{ab} a_p^{\dagger}a_q^{\dagger}a_ba_a + ... \nonumber \\
S_v = S_{1v} + S_{2v} + ...\nonumber \\
= \sum_{p \ne v} s_v^p a_p^{\dagger}a_v + \frac {1}{2}\sum_{bpq} s^{pq}_{vb} a_p^{\dagger}a_q^{\dagger}a_ba_v + ...
\end{eqnarray}
and $C(1,1), C(1,2), ... etc. $ are the perturbation co-efficients for each order of the corrected wavefunction. The first index of the superscript represents the order of the pertubation and the second represents the excitation level. Therefore, each T- operator accounts for the correlation effects from all orders of the perturbed wavefuction. The above relation shows that unlike in the MBPT, where the correction to correlation is computed order by order,
in coupled-cluster theory once the leading correlation corrections are
identified, a subset of terms to all orders which improve these corrections
are also included. Diagrammatic representation for this theory has been shown in the Fig. 1. \\ 

\begin{figure*}
\includegraphics[width=14.0cm]{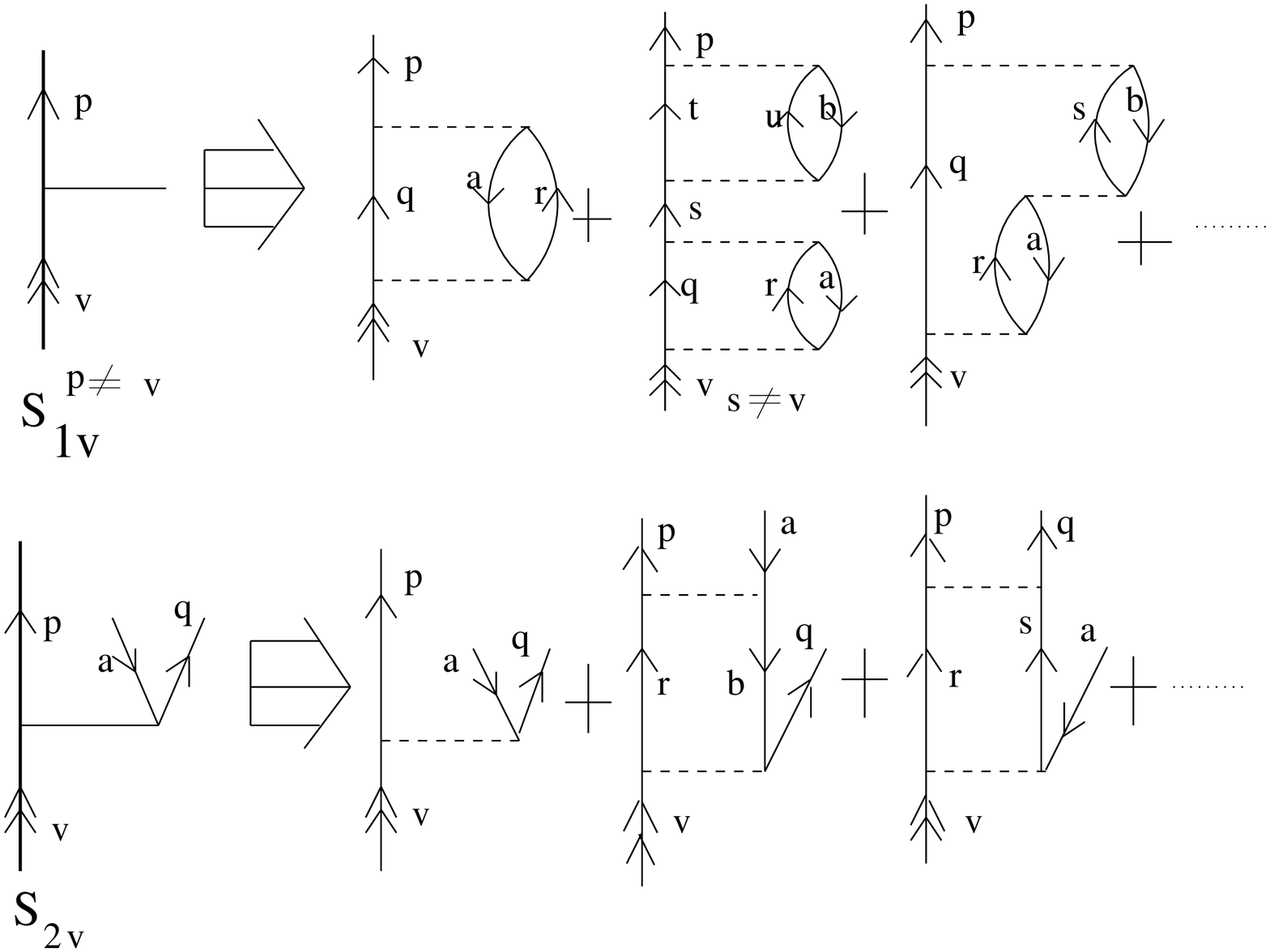}
\caption{Factorisation of all order CC amplitude diagrams for $S_{1v}$ and $S_{2
v}$ operators in terms of counterpart MBPT diagrams}
\end{figure*}

\noindent
The equations for the open-shell cluster amplitudes are determined from\cite{lindgren01,debasish}
\begin{eqnarray}
\langle \Phi_v|\bar{H_N}\{1+S_v\}|\Phi_v\rangle = \Delta E(v)  
\end{eqnarray}
and
\begin{eqnarray}
\langle \Phi_v^{*}|\bar{H_N}\{1+S_v\}|\Phi_v\rangle = -\Delta E(v) \langle \Phi_v^{*}|
\{S_v\}|\Phi_v\rangle
\end{eqnarray}
where, $\Delta E(v)$ is the electon attachment energy which is equal to the negative of the ionisation potential for the valence electron, {\it v}.\\

If 
$$ | \Psi_{v} \rangle = e^T \{ 1 + S_{v}\} |\Phi_{v}\rangle$$
and
$$ | \Psi_{v'} \rangle = e^T \{ 1 + S_{v'}\} |\Phi_{v'}\rangle$$
represent the ground and excited states with energies $E(v)$ and $E(v')$ respectively then the excitation energy is given by
\begin{equation}
E(v') - E(v) =  \Delta E(v') - \Delta E(v)
\end{equation}
\vskip1cm

\noindent
The transition matrix element for a general one particle operator can be expressed in coupled-cluster theory as
\begin{eqnarray}
 O_{fi}  &=& \frac {\langle\Psi_f | O | \Psi_i \rangle} {\sqrt{\langle\Psi_f|\Psi_f\rangle \langle\Psi_i|\Psi_i \rangle}} \nonumber \\
 &=& \frac {\langle \Psi_f | O | \Psi_i \rangle } {\sqrt{(1+N_f)(1+N_i)}} \nonumber \\
 &=& \frac {\langle \Phi_f | \{1+S_f^{\dagger}\} e^{T^{\dagger}} O e^T \{1 + S_i\} | \Phi_i\rangle } {\sqrt{(1+N_f)(1+N_i)}}\\
 &=& \frac {\langle \Phi_f | \{1+S_f^{\dagger}\} \bar O \{1 + S_i\} | \Phi_i\rangle } {\sqrt{(1+N_f)(1+N_i)}}
\end{eqnarray}
\noindent
where the normalisation terms for the {\it v$^{th}$} orbital is obtained from
\noindent
\begin{eqnarray}
N_v &=& \langle \Phi_v | S_v^{\dagger} [e^{T^{\dagger}} e^T] + S_v^{\dagger} [e^{T^{\dagger}} e^T] S_v + [e^{T^{\dagger}} e^T] S_v | \Phi_v\rangle\nonumber \\
 &=& \langle \Phi_v | S_v^{\dagger} \bar n_v + S_v^{\dagger} \bar n_v S_v + \bar n_v S_v^{\dagger} | \Phi_v\rangle
\end{eqnarray}
with
\noindent
\begin{equation}
\bar O=(e^{T^{\dagger}} O e^T)_{f.c.} + (e^{T^{\dagger}} O e^T)_{o.b.} + (e^{T^{\dagger}} O e^T)_{t.b.} + ....
\end{equation}
and
\noindent
\begin{equation}
\bar n_v = (e^{T^{\dagger}} e^T)_{f.c.} + (e^{T^{\dagger}} e^T)_{o.b.} + (e^{T^{\dagger}} e^T)_{t.b.} + ...
\end{equation}

\noindent
The f.c., o.b., t.b.,..etc abbreviations are used for the fully contracted, effective one-body, effective two-body ...etc terms respectively\cite{geetha02}. Terms containing only upto effective three-body diagrams will contribute to both the numerator and the denominator. The fully contracted terms are excluded on the basis of the linked-Diagram theorem\cite{lindgren01} in the evaluation of the $\bar O$ and $\bar N$. All the one-body terms have been taken into account as their contribution to the correlation effects is the largest. The dominat parts of the two-body terms have also been computed\cite{sahoo02,geetha02}. Finally, these terms are contracted with $S_f^{\dagger}$ and $S_i$ operators.\\ 

\noindent
Contributions from the normalisation factor have been determined in the following way 
\begin{eqnarray}
Norm = \langle \Psi_f | O | \Psi_i \rangle \{ \frac {1}{\sqrt{(1+N_f)(1+N_i)}} - 1 \}
\end{eqnarray}

\section{Results and Discissions:}

\begin{table*}
\begin{ruledtabular}
\begin{center}
\caption{Ionization potential energies of different states of $Pb^+$}
\begin{tabular}{lccccc}
\hline
\hline
States  &  Koopman(cm$^{-1}$) & MBPT(2)(cm$^{-1}$) &CCSD(T)(cm$^{-1}$) & Expt.(cm$^{-1}$) & \% of accuracy\\ 
\hline
$6p_{1/2}$ & 114015 & 121898 & 120126 & 121208 & 0.8 \\   
$6p_{3/2}$ & 100402 & 108041 & 106416 & 107123 & 0.6 \\
\hline
\hline
\end{tabular}
\end{center}
\end{ruledtabular}
\end{table*}

\noindent
We have used Gaussian type orbitals(GTO) for the construction of single paricle orbitals of the Dirac-Fock wavefunction($|\Phi \rangle$), whose expression is given by\cite{rajat1} 
\begin{equation}
F_{i,k}^{(L/S)}(r) = \sum_i c_i^{(L/S)} r^k e^{-\alpha_i r^2}
\end{equation}
with k=0,1,2,... for $s$, $p$, $d$,$\cdots$ respectively. The function $F_{i,k}^{(L/S)}(r)$ stands for the large(L) and small(S) components of the dirac wavefunction. $ c_i^{(L/S)}$ is the expansion coefficient of the corresponding large and small components respectively. The kinetic balance condition\cite{stanton01} has been imposed between the large and small components of the GTOs. For the 
exponents, the even tempering condition
\begin{equation}
\alpha_i = \alpha_{i-1} \beta , \hspace*{2cm} i=1,\cdots,N
\end{equation}
has been applied. Here, $N$ stands for the total number of basis functions for a specific symmetry.
In the present calculation, we have taken $\alpha_{\circ}$ = 0.00825 and $\beta$ = 2.73 for all symmetries. We have considered 13$s_{1/2}$, 13$p_{1/2}$, 13$p_{3/2}$, 11$d_{3/2}$, 11$d_{5/2}$, 8$f_{5/2}$, 8$f_{7/2}$, 7$g_{7/2}$ and 7$g_{9/2}$ active orbitals. All core electrons have been excited in the present calculation.
\\

\noindent
We have obtained  an accuracy of better than one percent for the ionisation potentials of both $6 ^2P_{1/2}$ and $6 ^2P_{3/2}$ states. The ionisation potential at the Dirac-Fock level is calculated using Koopman's theorem. As can be seen from table I the correlation contributions are about 5\% and 6\% for the $6 ^2P_{1/2}$ and the $6 ^2P_{3/2}$ states respectively. However, the CC excitation energy improves only by about one and half percent over the DF value and deviates by 2.7\% from the experimental value. The excitation energy between these two states is calculated using eq. (3.18) and it's value is given in table IV.   
From table II, it is clear that for both the M1 and E2 transition matrix elements, the total contribution of $OS_{1i}$ and its conjugate is larger than $OS_{2i}$ and its conjugate term. In its lowest order $OS_{1i}$ corresponds to the Brueckner pair correlation and $OS_{2i}$ to core polarisation. The largest contribution to electron correlation comes from the pair correlation effects. The reduced transition matrix elements for M1 and E2 operators are given in table II. The important contributions from the one-body terms of these two quantities are given in table III.\\
 
\begin{table*}
\begin{ruledtabular}
\begin{center}
\caption{Contributions from the important terms to the M1 and E2 transition matrix elements in a.u..}
\vspace*{0.1cm}

\item{(i)} Contribution from one-body terms:
\vspace*{0.1cm}

\begin{tabular}{lcc}
\hline
\hline
Terms  & Contributions for M1 & Contributions for E2 \\
\hline\\
$\bar O$ & -1.12021925042 & 9.05230735868 \\
$\bar O  S_{1i}$ &  -0.00517412601 & -0.18227019838 \\
$S_{1f}^{\dagger} \bar O$ & 0.00553162299 & -0.25852998602 \\
$\bar O  S_{2i}$ & 0.00017425249 & -0.09056241322 \\
$S_{2f}^{\dagger} \bar O$ & -0.00031546156 & 0.08833942426 \\
$S_{1f}^{\dagger} \bar O S_{1i}$ & -0.00112434477 & 0.01932259992 \\
$S_{2f}^{\dagger} \bar O S_{2i}$ & -0.02963327056  & 0.19544605525 \\
Norm. & 0.00989789079 & -.07635041195 \\
\hline\\
\end{tabular}
\vspace*{0.1cm}

\item{(ii)} Contribution from two-body terms:
\vspace*{0.1cm}

\begin{tabular}{lcc}
\hline\\
Terms & Contributions for M1 & Contributions for E2\\
\hline\\
$T_1^{\dagger} O S_{2i}$ & 0.00186337223 & -0.01235825671 \\
$S_{2f}^{\dagger} O T_1$ & 0.00188968606 & -0.01257926784 \\
$ T_2^{\dagger} O S_{2i}$ & -0.00015858582 & 0.00111460180 \\
$S_{2f}^{\dagger} O T_2$ & 0.00011723313 & 0.00049241644 \\
\hline
Total & -1.13709985958 & 8.77139789789 \\
\hline
\hline
\end{tabular}
\end{center}
\end{ruledtabular}
\end{table*}

\begin{table*}
\begin{ruledtabular}
\begin{center}
\caption{Important contributions from the individual one-body terms in a.u.:}
\begin{tabular}{lcc}
\hline
Terms & Contributions for M1 & Contributions for E2\\
\hline
Dirac-Fock & -1.137290674575 & 9.21109641205 \\
$O T_{1}$ & 0.00022055250 &  0.00094567191 \\
$T_{1}^{\dagger} O$ & -0.00024161331 & 0.00173923377 \\
$O S_{1i}$ & -0.00553158375 & -0.18040253154 \\
$S_{1f}^{\dagger} O$ &  0.00530735776 & -0.25744792175 \\
$S_{1f}^{\dagger} O S_{1i}$ & -0.00451820627 & 0.01933124629 \\
\hline
\hline
\hline
\end{tabular}
\end{center}
\end{ruledtabular}
\end{table*}

\begin{table*}
\begin{ruledtabular}
\caption{Lifetime of $6 ^2P_{3/2}$-state in sec.}
\begin{tabular}{|lccccc|}
\hline
\hline
& Excitation Energy & Wavelength  & M1- Channel & E2- Channel & Total\\ 
& (cm$^{-1}$) & (in \AA) & ($\tau_1$) & ($\tau_2$)  & ($\tau$) \\
\hline
Dirac-Fock & 13613 & 7346  & 0.0455 & 0.909 & 0.0433 \\
MBPT(2) & 13857 & 7216 & 0.0436 & 1.03 & 0.0418 \\
CCSD(T) & 13710 & 7294 & 0.0445 & 0.958 & 0.0425 \\
Expt\cite{werth02} & 14085 & 7100  &  &  & 0.0412(7) \\ \hline
\hline
\hline
\hline
\end{tabular}
\end{ruledtabular}
\end{table*}

\noindent
The net probability for a given transition which allows two different channels is given by\cite{drake}
\begin{equation}
A = A_1 + A_2
\end{equation}
This can be expressed as
\begin{equation}
\frac {1}{\tau} = \frac {1}{\tau_1} + \frac {1}{\tau_2}
\end{equation}
where $\tau_1$ and $\tau_2$ are the lifetimes through different branches and $\tau$ is the total lifetime. The results obtained for $'\tau_1'$ which represents the M1 lifetime and $'\tau_2'$ the E2 lifetime are given in table IV. The lifetime for the $6 ^2P_{3/2}$ state obtained using RCC is 0.0425 second which has an accuracy about 3\%. The lifetime result of our second order RMBPT calculation is 0.0418 second. This suggests that there would be a strong cancellation of the correlation effects from the higher order RCC contributions.

\section{Conclusion}
\noindent
Relativistic coupled-cluster theory has been applied to calculate ionisation potential and transition matrix elements for M1 and E2 operators to determine the lifetime for the $6 ^2P_{3/2}$ state of the singly ionised lead. The accuracy of the calculation is about 3\%.

\section{Acknowledgment}
\noindent
We would like to thank Prof. G\"unther Werth for his valuable discussions and suggesting for the work reported in this paper. DM thanks the CSIR(New Delhi) for the research grant no. 01(1624)/EMR-II. The computations were carried out on the Teraflop Supercomputer, C-DAC, Bangalore, India.

\end{document}